\documentclass[amsmath,amssymb]{nature}
\usepackage{bm}
\usepackage{amsmath}
\usepackage{graphicx}
\usepackage{color}
\usepackage{upgreek}
\usepackage[displaymath, mathlines]{lineno}
\newenvironment{minilinespace}{
\baselineskip = 1mm
}

\title{\small \rm {Nature Communications 7, 13039 (2016),\\
doi:10.1038/ncomms13039.\\}
\\
\Large \bf{Magnetodielectric detection of magnetic quadrupole order in Ba(TiO)Cu$_4$(PO$_4$)$_4$ with Cu$_4$O$_{12}$ square cupolas}}

\author{K. Kimura$^{1\ast}$, P. Babkevich$^2$, M. Sera$^{1}$, M. Toyoda$^{3}$, K. Yamauchi$^{4}$,
G.S. Tucker$^{2,5}$, J. Martius$^2$, T. Fennell$^5$, P. Manuel$^6$, D.D. Khalyavin$^6$, R.D. Johnson$^6$,
T. Nakano$^{7}$, Y. Nozue$^{7}$, H.M. R\o{}nnow$^{2}$ \&  T. Kimura$^{1}$}

\begin{document}

\maketitle

\begin{affiliations}
\item Division of Materials Physics, Graduate School of Engineering Science, Osaka University, Toyonaka, Osaka 560-8531, Japan
\item Laboratory for Quantum Magnetism, Institute of Physics, \'Ecole Polytechnique F\'ed\'erale de Lausanne (EPFL), CH-1015 Lausanne, Switzerland
\item Department of Physics, Tokyo Institute of Technology, Meguro-ku, Tokyo 152-8550, Japan
\item ISIR-SANKEN, Osaka University, Ibaraki, Osaka 567-0047, Japan
\item Laboratory for Neutron Scattering and Imaging, Paul Scherrer Institut, CH-5232 Villigen, Switzerland
\item ISIS facility, STFC Rutherford Appleton Laboratory, Chilton, Didcot, Oxfordshire, OX11 0QX, United Kingdom
\item Graduate School of Science, Osaka University, Toyonaka, Osaka 560-0043, Japan
\end{affiliations}

\begin{minilinespace}
\begin{flushleft}
\noindent{$^\ast$To whom correspondence should be addressed. E-mail: kentakimura@mp.es.osaka-u.ac.jp}\\
\normalsize{\ \ \ \ \ \ \ \ \ }
\end{flushleft}
\end{minilinespace}

\begin{abstract}
%\linenumbers
In vortex-like spin arrangements, multiple spins can combine into emergent multipole moments. Such multipole moments have broken space-inversion and time-reversal symmetries, and can therefore exhibit linear magnetoelectric (ME) activity.
Three types of such multipole moments are known: toroidal, monopole, and quadrupole moments. So far, however, the ME-activity of these multipole moments has only been established experimentally for the toroidal moment. Here, we propose a magnetic square cupola cluster, in which four corner-sharing square-coordinated metal-ligand fragments form a noncoplanar buckled structure, as a promising structural unit that carries an ME-active multipole moment. We substantiate this idea by observing clear magnetodielectric signals associated with an antiferroic ME-active magnetic quadrupole order in the real material Ba(TiO)Cu$_4$(PO$_4$)$_4$. The present result serves as a useful guide for exploring and designing new ME-active materials based on vortex-like spin arrangements.
\end{abstract}

\begin{figure}[htbp]
%\internallinenumbers
\begin{center}
\hspace{0cm}
\includegraphics[width=0.9\textwidth]{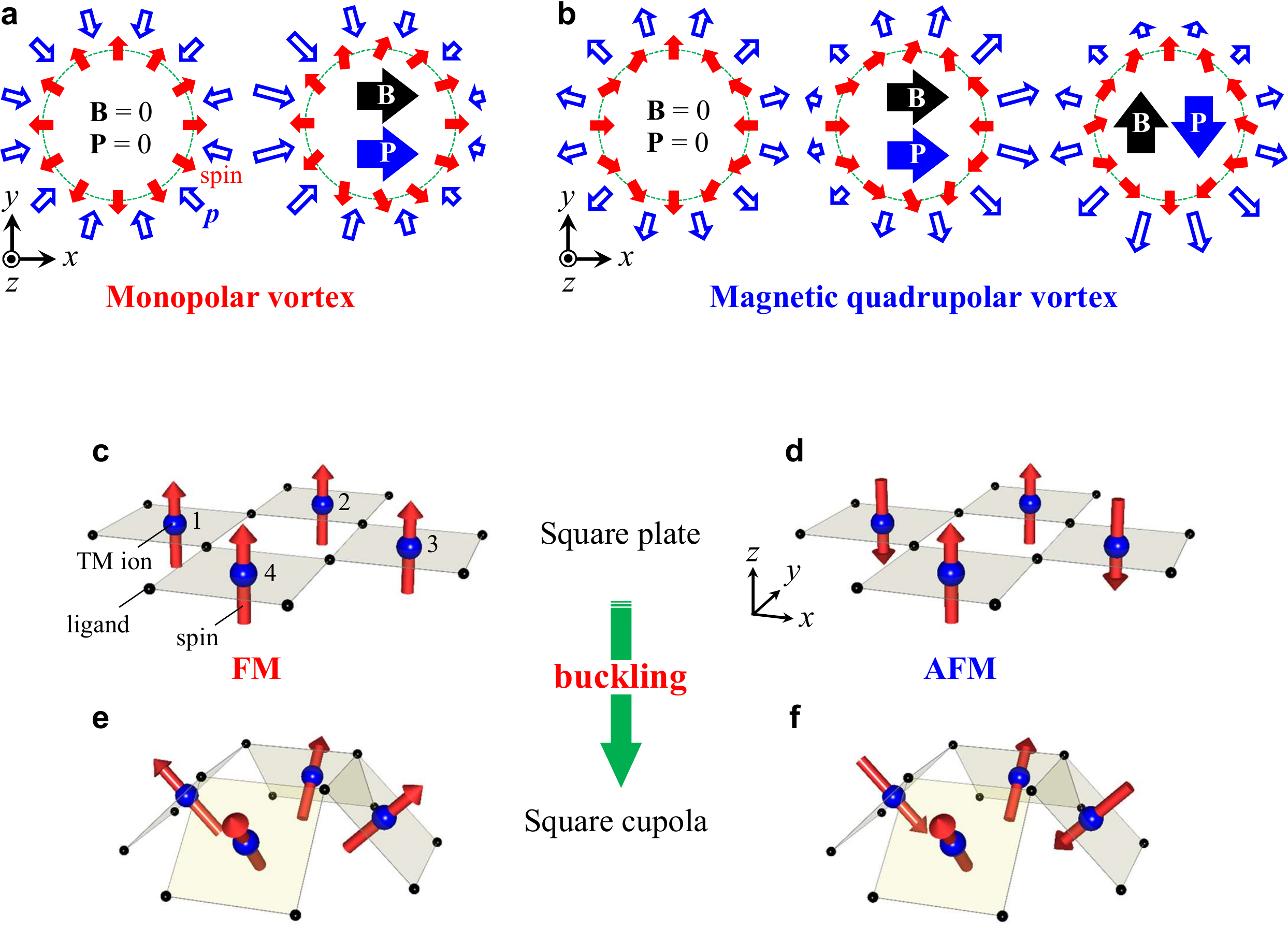}
\caption {
{\bf Conceptual design route of ME-active spin cluster.}
{\rm
{\bf a},{\bf b}, Schematic illustrations of monopolar ({\bf a}) and magnetic quadrupolar ({\bf b}) vortices in the absence and presence of a magnetic field $B$ (solid black arrow). The red solid, blue open, and blue solid arrows represent the orientation of spin, the local electric polarization $p$, and macroscopic electric polarization $P$, respectively.
{\bf c},{\bf d}, A square planar cluster composed of ferromagnetically (FM) ({\bf c}) and antiferromagnetically (AFM) ({\bf d}) coupled four spin plaquette, respectively. Blue and black balls represent the transition-metal (TM) ions ($n=1\sim4$) and ligands, respectively.
{\bf e},{\bf f}, Square cupola clusters after buckling the cluster in ({\bf c}) and ({\bf d}), respectively. The FM square cupola cluster in ({\bf e}) has the monopole moment $(\displaystyle a =\sum_{n=1\sim4}\bf r_{\it n} \cdot  \bf S_{\it n})$ while the AFM square cupola cluster in ({\bf f}) has the magnetic quadrupole moment $(\displaystyle q_{x2-y2}=\sum_{n=1,3}\bf r_{\it n} \cdot  \bf S_{\it n}-\sum_{\it n'=\rm{2,4}}\bf r_{\it n'} \cdot  \bf S_{\it n'})$, where $\bf r_{\it n}$ is a position vector of the TM ion $n$ from the centre of the four TM ions and $\bf S_{\it n}$ is a spin moment.
}
\label{fig1}
}
\end{center}
\end{figure}

%\linenumbers
In magnetic materials, noncollinear spin arrangements often emerge when spins are placed in a particular lattice geometry such as geometrically frustrated lattices\cite{Greedan2001}. Symmetry-breaking properties of an additional multispin degree of freedom inherent in the noncollinearity can generate various anomalous magnetic phenomena. A well-known example is spiral-spin-driven ferroelectricity arising from vector spin chirality with broken space-inversion symmetry\cite{Kimura2003,Katsura2005,Mostovoy2006,Sergienko2006}. Another example is the magnetoelectric (ME) effect -- magnetic field ($B$) control of electric polarization ($P$) and electric field ($E$) control of magnetization ($M$) -- which originates from magnetic multipole moments that break both space-inversion and time-reversal symmetries\cite{Astrov1960,Schmid2001,Fiebig2005,VanAken2007,Spaldin2008,Yamaguchi2013,Zimmermann2014}.
Three types of symmetrically distinct ME-active multipole moments are known:
the toroidal moment (${\bf t} \propto \Sigma_{n}{\bf r}_{n}\times {\bf S}_{n}$), the monopole moment ($a \propto \Sigma_{n}{\bf r}_{n} \cdot {\bf S}_{n}$), and the magnetic quadrupole moment
($q_{ij}\propto \Sigma_{n}[r_{ni} S_{nj} + r_{nj} S_{ni} - \frac{2}{3} \delta_{ij} {\bf r}_{n} \cdot {\bf S}_{n}$]), where $n$ represents a label of the spin ${\bf S}_{n}$ at the position vector ${\bf r}_{n}$ and $i$, $j$ denote the $x$, $y$, or $z$ axis\cite{Dubovik1990,Schmid2001,Spaldin2008,Spaldin2013}.
These quantities can be finite in a specific vortex-like spin arrangement. For example, the spin arrangements illustrated in  Figs.~\ref{fig1}a and \ref{fig1}b have the monopole moment $a$ and the quadrupole moment $q_{x2-y2} (=q_{xx}-q_{yy})$, respectively, which therefore allow for the linear ME effect determined by the corresponding ME tensor (for more details, see Supplementary Note 1).
Previous theoretical studies\cite{Mostovoy2006,Delaney2009} predict that a $B$-induced vortex deformation can generate finite $P$ in toroidal and monopolar vortices (Fig.~\ref{fig1}a), and also in the quadrupolar vortex as shown in (Fig.~\ref{fig1}b), which is independent of the ME coupling mechanism.
Of the three types of multipole moments, the uniform ordering of toroidal moments (ferrotoroidicity) and associated ME activities have been experimentally probed via various techniques\cite{Popov1998,VanAken2007,Ressouche2010,Baum2013,Zimmermann2014,Toledano2015}. In particular, an observation of ferrotoroidic domain structure\cite{VanAken2007} and its hysteretic switching by crossed magnetic and electric fields\cite{Baum2013,Zimmermann2014} have led researchers propose ferrotoroidicity to be a fourth form of primary ferroic orders in the fundamental scheme based on the order-parameter symmetries with respect to the space-inversion and time-reversal operations. The ME-activity of the other magnetic multipole moments, however, has never been established experimentally. Here, we propose a strategy for the realization of these ME-active multipole moments in a material starting from a simple magnetic cluster.

The magnetic cluster considered here is illustrated in Figs.~\ref{fig1}c and \ref{fig1}d. It consists of four transition-metal ions forming a spin plaquette and twelve coplanar ligands such as oxygen. This is a fragment of a very common lattice seen in many inorganic materials including the high-$T_{c}$ cuprate superconductors\cite{Tokura1990} and infinite-layer iron oxides\cite{Tsujimoto2007}. If we introduce a single-ion anisotropy (or exchange anisotropy) normal to the metal-ligand plane and simple ferromagnetic (FM) or antiferromagnetic (AFM) interactions between the spins, the FM and AFM spin-plaquette show an all-up (or all-down) and an up-down-up-down structure, respectively. Both are ME-inactive, because they have neither toroidal, monopole, nor quadrupole moments. To induce ME-activity, we introduce a buckling deformation that transforms the cluster geometry to one of the Johnson solids\cite{Johnson1966} known as square cupola (Figs.~\ref{fig1}e and \ref{fig1}f), while assuming that the anisotropy and exchange interactions remain unchanged. With this buckling, the FM spin plaquette has monopole components in addition to ferromagnetic components, while the AFM spin plaquette has magnetic quadrupole components. Therefore, in principle, both square cupola clusters should exhibit ME-activity.

\begin{figure}
\begin{center}
\hspace{0cm}
\includegraphics[width=1\textwidth]{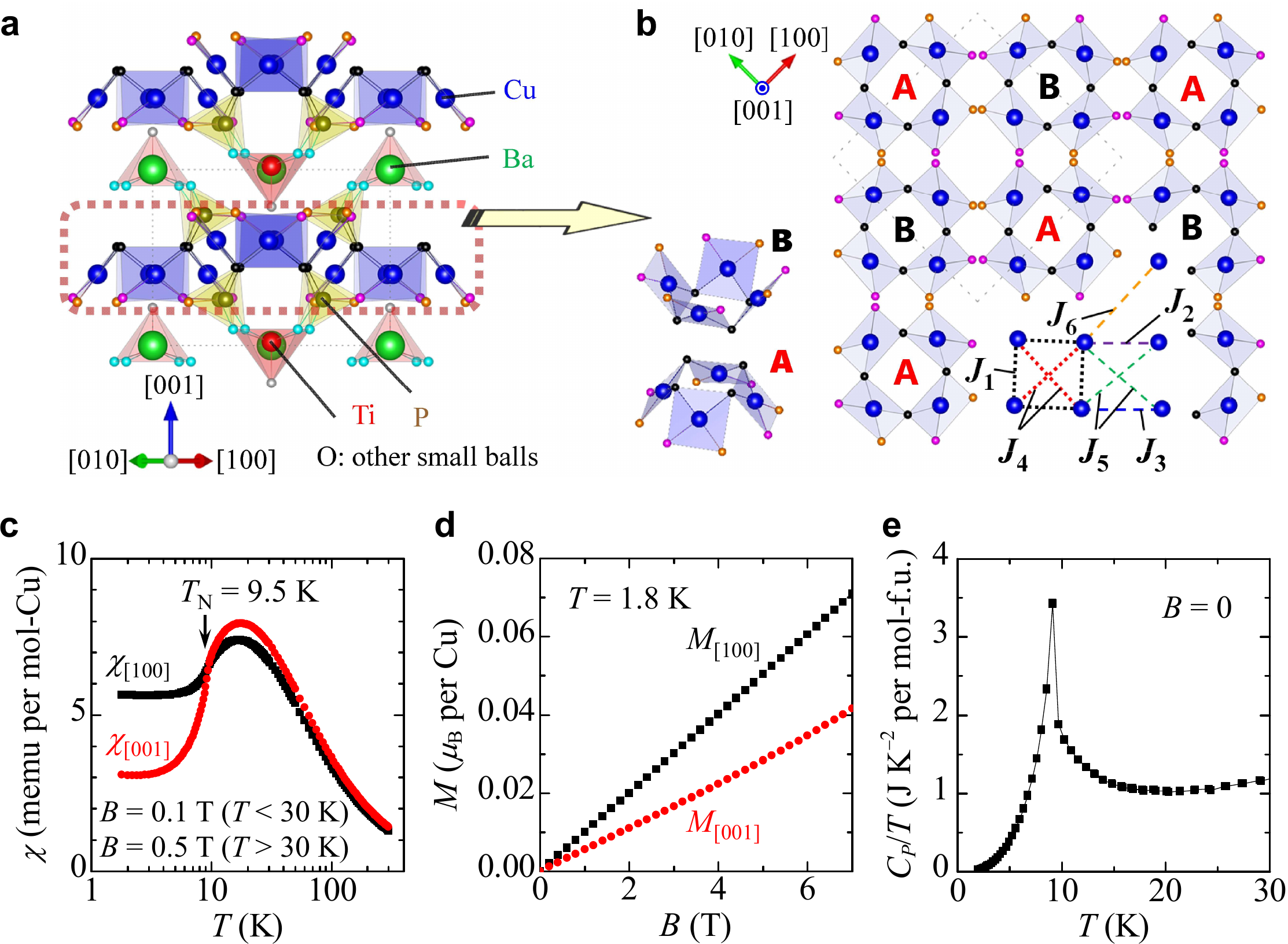}
%\internallinenumbers
\caption {
{\bf Crystal structure and magnetic properties of Ba(TiO)Cu$_4$(PO$_4$)$_4$.}
{\rm
{\bf a}, Crystal structure of  Ba(TiO)Cu$_4$(PO$_4$)$_4$ viewed along the [110] direction. The gray dotted line represents a unit cell.
{\bf b}, Square cupola layer viewed along the [001] direction. Two types of Cu$_4$O$_{12}$ square cupola clusters, A (upward) and B (downward), are shown. Intraplaquette ($J_{1}$ and $J_{4}$) and interplaquette ($J_{2}$, $J_{3}$, $J_{5}$, and $J_{6}$) exchange interactions are denoted by dotted and dashed lines, respectively.
{\bf c}, Temperature dependence of magnetic susceptibility for the field along the [100] (black square) and [001] (red circle) directions.
{\bf d}, Magnetization curves at 1.8 K.
{\bf e}, Temperature dependence of specific heat divided by temperature.
}
\label{crystalmag}
 }
\end{center}
\end{figure}

To experimentally verify this ME-design strategy, we have searched for a real material that comprises magnetic square cupola clusters. \makeatletter\renewcommand{\@cite}[2]{#1}\makeatother
The candidate compound found is Ba(TiO)Cu$_4$(PO$_4$)$_4$, a recently synthesized magnetic insulator crystallizing in a chiral tetragonal structure with space group $P42_{1}2$ (ref.~\cite{KKimura2016}).
\makeatletter\renewcommand{\@cite}[2]{\leavevmode%
\hbox{$^{\mbox{\the\scriptfont0 #1}}$}}\makeatother
The crystal structure is illustrated in Figs.~\ref{crystalmag}a and \ref{crystalmag}b. The magnetic properties are dominated by  Cu$^{2+}$ ions ($S = 1/2$) with square planar coordination of oxygen ions. Notably, the crystal structure comprises an irregular Cu$_4$O$_{12}$ square cupola cluster formed by four corner-sharing CuO$_4$ planes -- i.e., an experimental realization of Figs.~\ref{fig1}e or \ref{fig1}f. Two types of square cupola clusters, A (upward) and B (downward), distinguished by their direction with respect to the $c$-axis, align alternatingly in the $ab$-plane to form a layered structure which we call a square cupola layer. Importantly, the nature of the square cupola is expected to be preserved in this compound because neighboring square cupola clusters do not share any oxygen, suggesting weak inter-cluster couplings (Fig.~\ref{crystalmag}b). These features make this material suitable for testing our ME-design strategy. Therefore, we have studied magnetic and ME properties of Ba(TiO)Cu$_4$(PO$_4$)$_4$, demonstrating that the magnetic structure is describable by an antiferroic magnetic quadrupole order and, moreover, an associated ME-activity is manifested in the magnetodielectric properties. These results successfully verify our ME-design strategy.

\noindent{{\bf Results}}\\
{\bf Magnetic properties.}
 The temperature ($T$) dependence of magnetization ($M$) divided by $B$ ($\chi \equiv M/B)$ applied along the [100] and [001] axes is shown in Fig.~\ref{crystalmag}c. Fits to the high temperature data ($T > 100$ K) of $\chi_{[100]}$ and $\chi_{[001]}$ using the Curie-Weiss law yield effective moments of 1.92(1) $\mu_{\rm B}\rm{/Cu}$ and 1.96(1) $\mu_{\rm B}\rm{/Cu}$, respectively, which are typical for Cu$^{2+}$ ions, and AFM Weiss temperatures ($\theta_{\rm CW}$) of $-33.2(6)$ K and $-30.1(2)$ K. On cooling, a broad maximum appears at around 17 K, followed by a clear anomaly at $T_{\rm N} = 9.5$ K, below which $\chi$ shows an anisotropy. Because of the two dimensional (2D) nature of the square cupola layers due to the separation by nonmagnetic Ba and Ti layers, the broad maximum suggests a development of short-range correlation within each cluster and/or 2D layer. The anomaly at $T_{\rm N} = 9.5$ K indicates the onset of AFM long-range order due to weak inter-layer couplings. Notably, both $\chi_{[100]}$ and $\chi_{[001]}$ remain finite at the lowest temperature measured, which indicates that the magnetic structure is not a simple collinear antiferromagnet. No metamagnetic transition is observed up to $B=7$ T, as demonstrated by approximately linear magnetization curves at $T=1.8$ K (Fig.~\ref{crystalmag}d). The AFM transition is also evidenced by a peak in specific heat $C_P$, as shown in Fig.~\ref{crystalmag}e.

\begin{figure}
\begin{center}
\hspace{0cm}
\includegraphics[width=0.8\textwidth]{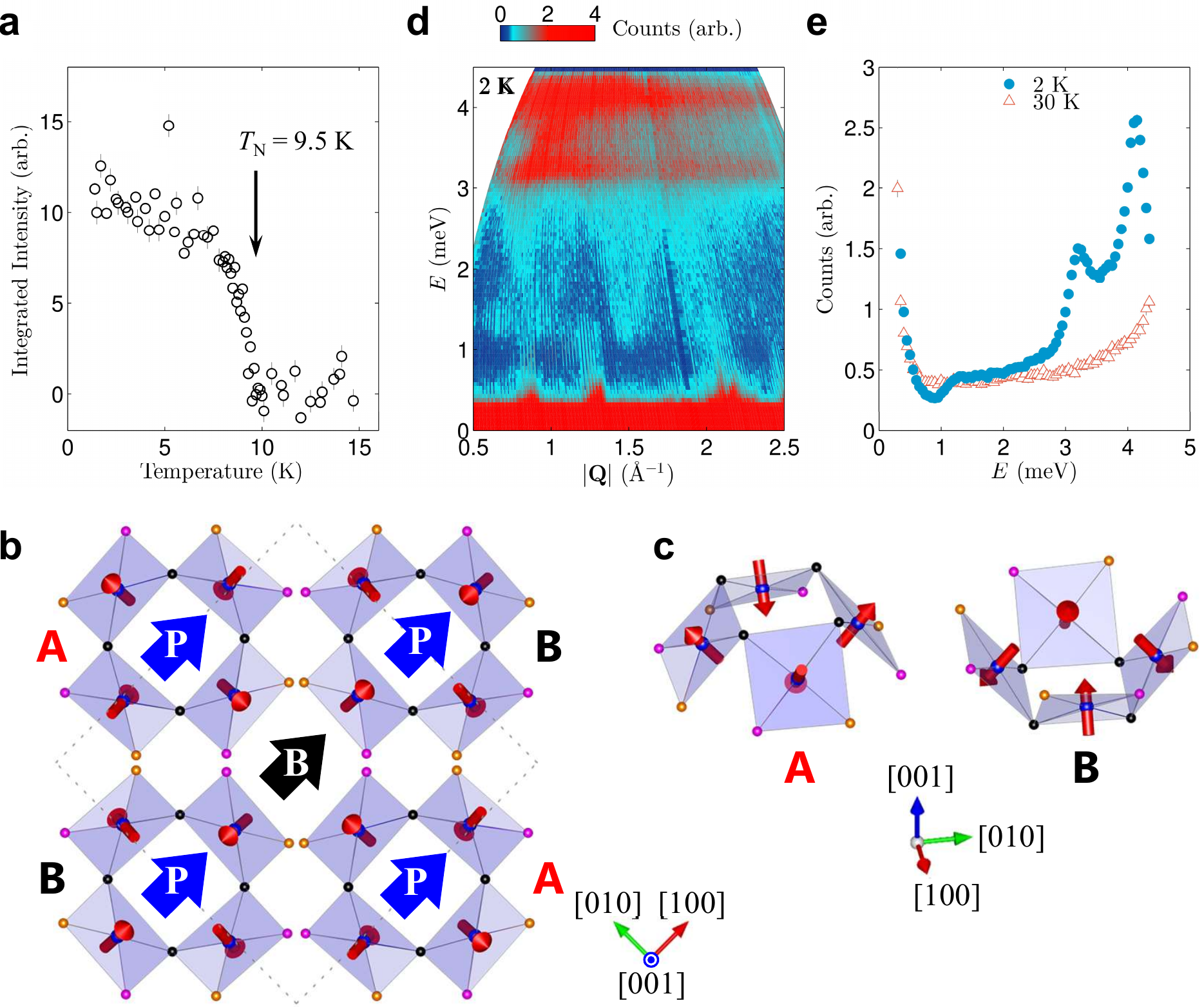}
%\internallinenumbers
\caption {
{\bf Magnetic structure and magnetic excitations probed through neutron scattering experiments.}
{\rm
{\bf a}, Temperature evolution of the sum of the integrated intensities of the $00\frac{1}{2}$,  $11\frac{1}{2}$, and $21\frac{1}{2}$ magnetic reflections (open circles).
{\bf b}, Magnetic structure of a single square cupola layer. Expected direction of electric polarization $P$ in each square cupola induced by a magnetic field ($B$) along the [100] direction is indicated.
{\bf c}, Detailed configuration of magnetic moments in square cupolas A and B.
{\bf d}, Inelastic neutron powder intensity map at 2 K as a function of momentum $|\mathbf{Q}|$ and energy $E$ transfer. Measurements were recorded from a powder sample using incident neutron energy of 6 meV.
{\bf e}, A constant-$Q$ cut integrated between $1.3 < |\mathbf{Q}| <1.6$ \AA$^{-1}$ at 2 and 30 K.
}
\label{neutron}
}
\end{center}
\end{figure}

For microscopic characterization of the magnetic properties, we employ neutron diffraction. As depicted in Fig.~\ref{neutron}a, below $9.5$ K we observe new Bragg peaks emerging corresponding to magnetic ordering of moments.
The magnetic reflections can be indexed using a single propagation wavevector $\mathbf{k} = (0,0,0.5)$, which corresponds to a doubling of the unit cell along the $[001]$ direction. Symmetry analysis indicates that the magnetic representation can be decomposed into $\Gamma_{\rm mag} = 3\Gamma_1 + 3\Gamma_2 + 3\Gamma_3 + 3\Gamma_4 + 6\Gamma^{(2)}_5$, where the irreducible representation (IR) $\Gamma^{(2)}_5$ is two-dimensional and the others are one-dimensional.
The best fit magnetic structure obtained by the Rietveld refinement with the magnetic $R$-factor of 11.5\% ($\Gamma_3$ IR) at 1.5 K is illustrated in Fig.~\ref{neutron}b for a single square cupola layer. The magnetic structure is noncollinear, with the moments tilted away from the $c$-axis in such a way that they are approximately perpendicular to the CuO$_4$ plane. Notably, in each Cu$_4$O$_{12}$ square cupola (Fig.~\ref{neutron}c), the $c$-axis components of magnetic moments align in the up-down-up-down manner while the $ab$-plane components rotate by 90 degrees. This demonstrates an experimental realization of the AFM square cupola considered in Fig.~\ref{fig1}b, which carries a magnetic quadrupole moment composed of an almost pure $q_{x2-y2}$ component. Moreover, the $q_{x2-y2}$ components in every Cu$_4$O$_{12}$ square cupola align uniformly, suggesting that the magnetic structure can be considered as a uniform order of mangnetic quadrupole moments within the $ab$-plane. Further details on this quadrupole order can be found in Supplementary Fig. 1 and Supplementary Note 2. An ordered moment of 0.80(3)$\mu_{\rm B}$ is found on each Cu site. The details of refinement are provided in Supplementary Figs. 2-4, Supplementary Tables 1 and 2, and Supplementary Note 3.

\noindent{{\bf Dominant exchange interactions.}}
The quadrupole-based description of the magnetic structure is valid if intraplaquette exchange interactions dominate over interplaquette ones. To examine this point, we have performed inelastic neutron scattering measurements.
Color plot of inelastic neutron spectra at $T = 2$ K is shown in Fig.~\ref{neutron}d. We observe two strong, flat bands of intensity close to 3.2 and 4.2 meV [see also Fig.~\ref{neutron}e]. This is consistent with the energy scale estimated from $\theta_{\rm CW} \approx -30$ K.
In addition, we find branches dispersing from magnetic zone centres with a gap energy of around 1 meV. On warming to 30 K, shown in Fig.~\ref{neutron}e, we find that both the gapped branches and the flat bands of intensity disappear as would be expected for excitations which are magnetic in origin. Further measurements using $E_{\rm i} = 12.1$ meV did not reveal any additional excitations.
The observed inelastic spectrum is consistent with strongly-coupled plaquettes with weak inter-plaquette interactions and an anisotropy, such as from Dzyaloshinskii-Moriya interaction which is symmetrically allowed in this material. The former would result in dispersive collective excitations and the latter create a spin-gap.
\makeatletter\renewcommand{\@cite}[2]{#1}\makeatother
Therefore, Ba(TiO)Cu$_4$(PO$_4$)$_4$ could potentially be an exciting system to examine the cross-over from local quantum levels to dispersive spin-waves of coupled spin clusters such as in Cu$_2$Te$_2$O$_5$(Cl,Br)$_2$ (ref.~\cite{Prsa2009}).
\makeatletter \renewcommand{\@cite}[2]{\leavevmode%
\hbox{$^{\mbox{\the\scriptfont0 #1}}$}}\makeatother

Further insights to exchange interactions are provided by density functional theory (DFT) calculations.
The electronic structure of Ba(TiO)Cu$_4$(PO$_4$)$_4$ was calculated by using GGA+$U$ method\cite{PBE1996,Liechtenstein1995}, where the on-site Coulomb replusion $U_\mathrm{eff}$ was set to 4 eV.
We have calculated the magnetic exchange coupling constants $J_k$ ($k = 1 \sim 6$) between the spins (see Fig.~\ref{crystalmag}b). Here, positive (negative) $J$ represents  FM (AFM) interactions.
We find that the strongest interaction is the nearest neighbor intraplaquette $J_1 = -3.0$ meV, followed in order by orthogonal interplaquette interaction $J_5 = -0.7$ meV and orthogonal intraplaquette $J_4 = -0.5$ meV. The other interactions are one order of magnitude smaller than $J_{1}$ ($J_2 = -0.2$ meV, $J_3 = 0.2$ meV, and $J_6 = -0.1$ meV).
From these coupling constants the Weiss temperature is calculated to be $-31$ K, which is comparable to the experimental value of $\theta_{\rm CW} \approx -30$ K.
We have also calculated $J_k$ with different values of $U_\mathrm{eff}$ (0 and 7 eV) and confirmed that the relative strengths of $J_k$ is almost insensitive to $U_\mathrm{eff}$.
Thus, our experimental and theoretical studies establish that the present material is a weakly-coupled plaquette antiferromagnet, which means that the quadrupole-based description of the magnetic structure is valid.
Furthermore, a preliminary spin-wave calculation\cite{Toth2015} using the exchange parameters ($J_1 \sim J_6$) obtained from DFT calculations gives a qualitative agreement to the measured inelastic neutron spectrum, which confirms the consistency between the experiments and the DFT results (see Supplementary Fig. 5). However, because of a relatively large number of exchange paths which must be taken into account, single crystal study is necessary for a detailed comparison of inelastic neutron spectrum and a model based on DFT parameters.

\noindent{{\bf Magnetodielectric effect.}}
The task is now to experimentally confirm the ME-activity of these quadrupole moments. Usually, it can be easily probed through the measurements of $B$-induced $P$ or $E$-induced $M$. In the present case, however, it is not straightforward because, as indicated by the $00\frac{1}{2}$ magnetic wave vector, the quadrupole moments are antiferroically coupled along the $c$-axis, which results in the cancellation of the associated linear ME-response.
Indeed, our high-precision pyroelectric current measurement ($<0.1$ pA) at a high-$B$ ($<9$ T) does not show any signal indicative of a macroscopic $B$-induced $P$ near $T_{\rm N}$. Nonetheless, we observe a clear signature for the ME-activity in the dielectric constant ($\varepsilon$), as shown in the following.

\begin{figure}
\begin{center}
\hspace{0cm}
\includegraphics[width=1\textwidth]{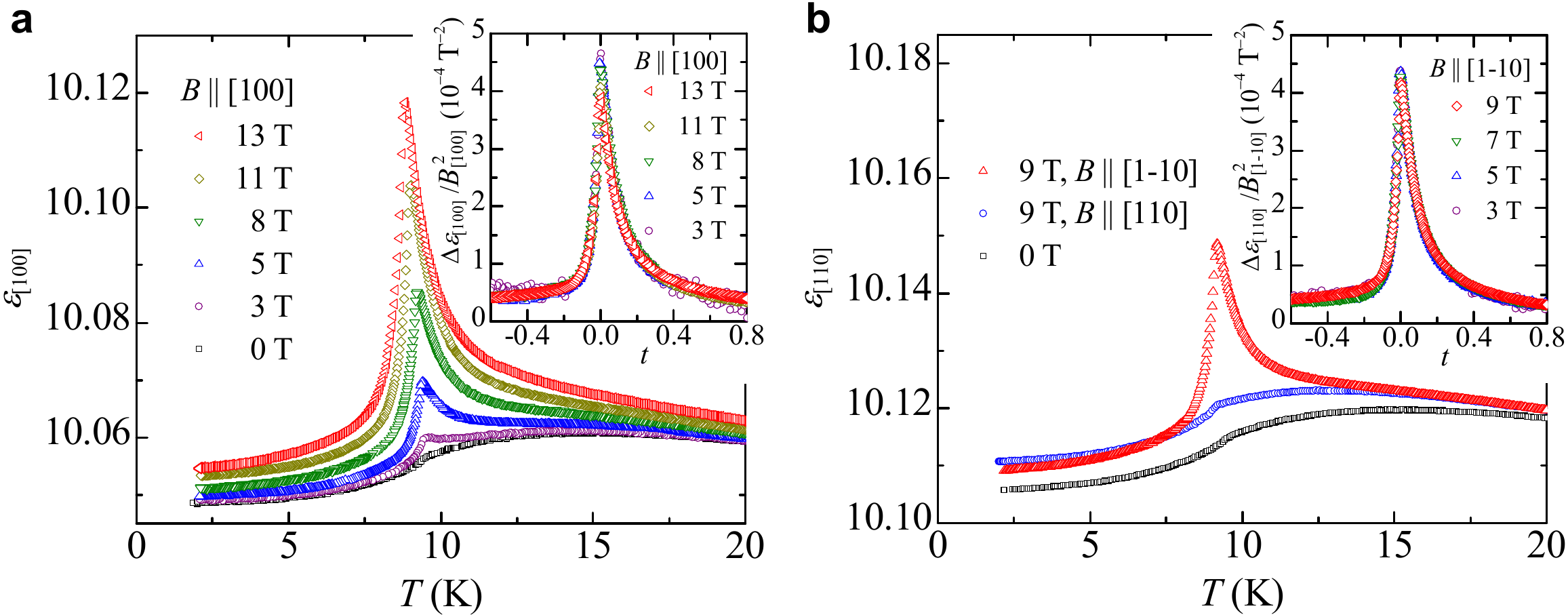}
%\internallinenumbers
\caption {
{\bf Magnetodielectric properties upon magnetic ordering.}
{\rm
{\bf a}, Temperature dependence of the dielectric constant along the [100] direction ($\varepsilon_{[100]}$) in various magnetic fields applied along the [100] direction ($B_{[100]}$). The inset shows the $B_{[100]}$-induced change ${\rm \Delta} \varepsilon_{[100]}(B_{[100]}) \equiv \varepsilon_{[100]}(B_{[100]})-\varepsilon_{[100]}(0~ \rm {T})$ divided by the square of $B_{[100]}$, ${\rm \Delta} \varepsilon_{[100]}/B_{[100]}^2$, as a function of a reduced temperature, $t \equiv (T-T_{\rm N})/T_{\rm N}$.
{\bf b}, Temperature dependence of $\varepsilon_{[110]}$ in $B=0$, $B_{[110]}=9$ T, and $B_{[1\mathchar`-10]}=9$ T. The inset shows the ${\rm \Delta} \varepsilon_{[110]}/B_{[1\mathchar`-10]}^2$ as a function of $t$.
}
\label{dielectric}
}
\end{center}
\end{figure}

Figure~\ref{dielectric}a shows the $T$-dependence of $\varepsilon$ along the [100] direction ($\varepsilon_{[100]}$) measured at selected $B$ applied along the [100] direction ($B_{[100]}$). While $\varepsilon_{[100]}$ shows only a slight anomaly at $T_{\rm N}$ in the absence of $B_{[100]}$, the application of $B_{[100]}$ induces a divergent peak toward $T_{\rm N}$.
\makeatletter\renewcommand{\@cite}[2]{#1}\makeatother
Together with the absence of $P$, this result suggests the onset of the $B$-induced antiferroelectric (AFE) order, which is similar to the $B$-induced ferroelectric order observed in linear ME materials such as Cr$_{2}$O$_{3}$ (ref.~\cite{Iyama2013}).
\makeatletter \renewcommand{\@cite}[2]{\leavevmode%
\hbox{$^{\mbox{\the\scriptfont0 #1}}$}}\makeatother
To quantitatively analyze the data, we define the $B_{[100]}$-induced component in $\varepsilon_{[100]}$ as ${\rm \Delta} \varepsilon_{[100]}(B_{[100]}) \equiv \varepsilon_{[100]}(B_{[100]})-\varepsilon_{[100]}(0)$.
The inset of Fig.~\ref{dielectric}a shows ${\rm \Delta} \varepsilon_{[100]}(B_{[100]})$ divided by the square of $B_{[100]}$,  ${\rm \Delta} \varepsilon_{[100]}/B_{[100]}^2$, as a function of a reduced temperature, $t \equiv (T-T_{\rm N})/T_{\rm N}$.
Strikingly, all the data approximately collapse onto a single curve, meaning that ${\rm \Delta} \varepsilon_{[100]}(B_{[100]})$ is proportional to the square of $B_{[100]}$. This scaling behavior of the $B$-induced divergent peak is characteristic in AFM linear ME materials such as MnTiO$_{3}$, which is accounted for on the basis of Landau free energy expansion involving the linear ME coupling term\cite{Mufti2011}.
No divergent behavior is observed for $\varepsilon_{[100]}$ in $B_{[010]}$ and $B_{[001]}$ and for $\varepsilon_{[001]}$ in all $B$ directions. (see Supplementary Fig. 7).

\noindent{{\bf Discussion}}\\
Because different ME-active multipole moments (${\bf t}$, $a$, and $q_{ij}$) lead to different forms of the ME tensor (Supplementary Fig. 6 and Supplementary Note 1), identifying the ME tensor in Ba(TiO)Cu$_4$(PO$_4$)$_4$ is crucial to verify that the observed ME response originates from the ME-activity of the quadrupole moments. To this end, we first discuss the observed ME response in terms of the linear ME effects in each square cupola layer. According to the transformation properties of  the present $\Gamma_{3}$ magnetic structure (Supplementary Table 2), the magnetic point group symmetry of each square cupola layer is $4'22'$, which breaks both the space-inversion and time-reversal symmetries and therefore allows for a linear ME effect given by the ME tensor\cite{Birss1966},
\begin{eqnarray}
\alpha_{\rm ME} = \left(
    \begin{array}{ccc}
      \alpha_{11} & 0 & 0 \\
      0 & -\alpha_{11} & 0 \\
     0 & 0 & 0
    \end{array}
  \right).
  \label{eqn1}
\end{eqnarray}
This ME tensor predicts that the application of $B_{[100]}$ induces $P$ along the [100] direction ($P_{[100]}$). In addition, the $00\frac{1}{2}$ magnetic wave vector indicates that the sign of the ME tensor of the neighboring layers is opposite to each other. This means that the system exhibits a $B_{[100]}$-induced AFE behavior along the [100] direction, consistent with the observed ME response. Note that this ME tensor allows, in principle, a macroscopic $q_{x2-y2}$ type quadrupole moment within the $ab$-plane (Supplementary Fig. 6), which is consistent with the uniform alignment of  $q_{x2-y2}$ quadrupole moments on the individual square cupolas.
Consistently, our group theory analysis in the framework of Landau theory of phase transitions\cite{Landau1980,Yamauchi2011,Yamauchi2014} shows that the quadrupole order can lead to the same ME effect as expected from the ME tensor in equation (1).
\makeatletter\renewcommand{\@cite}[2]{#1}\makeatother
Details of the analysis are provided in Supplementary Fig. 8, Supplementary Table 3 (refs.~\cite{Campbell2006},\cite{Aroyo2011},\cite{Aroyo2006},\cite{Aroyo2006b}), and Supplementary Note 4.
\makeatletter \renewcommand{\@cite}[2]{\leavevmode%
\hbox{$^{\mbox{\the\scriptfont0 #1}}$}}\makeatother

Next, we associate the observed result with the ME-activity of individual magnetic quadrupole moments.
As schematically illustrated in Fig.~\ref{fig1}b, while no net $P$ emerges from the magnetic quadrupole moment at $B=0$, the application of $B$ along the $x$- or $y$-axis induces $P$ parallel or antiparallel to the applied $B$, where the local principal axes parallel to the inward and outward spins are taken as the $x$- and $y$-axis, respectively. By applying this scenario to the single square cupola layer of the present system, we find that under $B_{[100]}$ all the quadrupole moments generate $P$ approximately parallel to the $B$ direction, resulting in macroscopic $P_{[100]}$ in the single square cupola layer (Fig.~\ref{neutron}b). Although a small diagonal component of local $P$ may be generated in each quadrupole moment due to the slight deviation of the inward spin direction from the [100] direction, this component cancels out in neighboring quadrupole moments because of the symmetry. The direction of $B$-induced $P$ is again consistent with that of the observed ME response and the ME tensor in equation (1). Note that there must be some contributions to the ME activity from the interplaquette interactions. However, the leading exchange interaction is the intraplaquette $J_{\rm 1}$, which provides a reasonable basis for the interpretation of the observed ME-activity in terms of the ME activity of individual square cupolas. Due to an AFM stacking of quadrupole moments along the [100] direction, the induced $P$ aligns antiferroelectrically along the [001] direction.

Furthermore, on the basis of the ME tensor given by equation (1) as well as the ME-activity of individual magnetic quadrupole moments (Fig.~\ref{fig1}b), one can expect that $P_{[110]}$ is induced by the application of ${B}_{[1\mathchar`-10]}$ but not by ${B}_{[110]}$, and the magnitude of ${B}_{[1\mathchar`-10]}$-induced $P_{[110]}$ is the same as that of ${B}_{[100]}$-induced $P_{[100]}$.
This expectation is experimentally demonstrated in the effects of ${B}_{[1\mathchar`-10]}$ and $B_{[110]}$ on $\varepsilon_{[110]}$ (Fig.~\ref{dielectric}b), together with the scaling behavior of $\varepsilon_{[110]}$ with the magnitude of ${\rm \Delta} \varepsilon_{[110]}/B^2_{[1\mathchar`-10]}$ comparable to that of ${\rm \Delta} \varepsilon_{[100]}/B_{[100]}^2$ (Fig.~\ref{dielectric}b, inset).
Thus, the observed AFE behavior can be nicely explained in terms of the ME-activity of quadrupole moments and their antiferroic alignment.

Finally, we discuss a possible microscopic origin for the observed ME-activity. The previous theoretical work demonstrated that a magnetostriction mechanism via non-relativistic superexchange interactions can generate large ME effects in monopolar and toroidal vortices\cite{Delaney2009}. Applying the similar discussion to the present quadrupolar vortex on a square cupola, we have found that the magnetostriction mechanism can also lead to the $B$-induced $P$, and the direction  of $B$ and $P$ is consistent with the experimental observation (see Supplementary Fig. 9). However, because other known mechanisms associated with a relativistic spin orbit interaction, e.g., the $d$-$p$ hybridization mechanism\cite{Arima2007} and the spin-current \cite{Katsura2005} (or inverse DM\cite{Sergienko2006}) mechanism, can also be possible. Therefore, more detailed understanding of the microscopic mechanism for the ME-activity is left for future study.

To conclude, we propose a magnetic square cupola cluster as a promising structural unit that carries the ME-active multipole moments, and confirm this idea by magnetodielectrically detecting the ME-activity of the magnetic quadrupole moments in a real material. This result indicates that ME-activity arising from magnetic multipoles can also be found in other square cupola based materials [for example, Na$_5$$A$Cu$_4$(AsO$_4$)$_4$Cl$_2$ ($A=$ Rb, Cs)\cite{Hwu2002} and
\makeatletter\renewcommand{\@cite}[2]{#1}\makeatother
[NH$_4$]Cu$_4$Cl(PO$_3$F)$_4$ (ref. \cite{Williams2015})].
\makeatletter \renewcommand{\@cite}[2]{\leavevmode%
\hbox{$^{\mbox{\the\scriptfont0 #1}}$}}\makeatother
In particular, the monopolar vortex, which can appear in the FM square cupola (Fig.~\ref{fig1}a), is interesting because it possesses a ferromagnetic component controllable by an electric field through the ME-coupling.
Another aspect that deserves attention in the present study is the discovery of the ME response due to the antiferroic ME-active multipole order. For example, this ME response allows for an $E$- or $B$-induced finite-$q$ magnetization or electric polarization, which might be useful for future nanoscale spintronics.
Moreover, the present discovery demonstrates that dielectric constant measurements can provide a rather straightforward, macroscopic signature for antiferroic order of any types of ME-active multipole moments. This is an important finding to explore a new state of matter such as a recently discussed antiferromonopolar state and antiferrotoroidic state\cite{Spaldin2013}.

\noindent{{\bf Methods}}\\
\noindent{{\bf Sample preparation and characetrization.}}
Single crystals of Ba(TiO)Cu$_4$(PO$_4$)$_4$ were grown by the flux method\cite{KKimura2016}. Powder X-ray diffraction (XRD) measurements on crushed single crystals confirmed a single phase. The crystals were oriented using Laue XRD. The crystal structures displayed in this article were drawn using VESTA software\cite{Momma2011}. Magnetization ($M$) measurements down to $T$ of 1.8 K and magnetic field ($B$) up to 7 T were performed using a commercial superconducting quantum interference device magnetometer (Quantum Design MPMS3). The specific heat ($C_{P}$) was measured down to 2 K by a thermal relaxation method using a commercial calorimeter (Quantum Design PPMS). For dielectric measurements, single crystals were cut into thin plates and subsequently silver electrodes were vacuum deposited on the pair of widest surfaces. The dielectric constant $\varepsilon$ was measured using an $LCR$ meter at an excitation frequency of 100 kHz. Pyroelectric current was measured by an electrometer (Keithley 6517) to monitor electric polarization.

\noindent{{\bf Neutron scattering experiments.}}
Neutron diffraction measurements were performed on a powder sample using the time-of-flight neutron diffractometer WISH at ISIS\cite{chapon-wish} and the DMC diffractometer at the SINQ spallation source\cite{Fisher}. Magnetic and nuclear structure refinements were performed using FullProf\cite{fullprof}. The inelastic neutron scattering measurements were carried out on a 17\,g powder sample using the time-of-flight spectrometer FOCUS at the SINQ spallation source\cite{janssen}. Using incident neutron energies of $E_{\rm i} = 6$ and 12.1 meV, the energy resolution in the energy transfer range of interest was approximately 0.2 and 0.7 meV, respectively.

\noindent{{\bf Density functional theory calculations.}}
Density functional theory (DFT) calculations were performed to estimate the magnitude of dominant magnetic interactions. The VASP (Vienna ab initio simulation package)\cite{Kresse1993,Kresse1994,Kresse1996,Kresse1996_cms} was used with a projector-augmented wave (PAW) basis set\cite{PAW,Kresse1999}.
The electronic exchange and correlation were described by the Perdew-Burke-Ernzerhof generalized gradient approximation (PBE-GGA)\cite{PBE1996}. The DFT$+U$ method\cite{Liechtenstein1995} was used for correction for strongly correlated Cu-$3d$ states.
We first calculated the electronic structure of Ba(TiO)Cu$_4$(PO$_4$)$_4$ with the experimental crystal structure. Then we obtained the total energy differences among several magnetic phases with different spin structures by performing spin-constrained DFT calculations. The magnetic exchange coupling constants $J_k$ were estimated as the best fit for the energy differences within an effective classical Heisenberg model. Our model Hamiltonian is defined as follows:
\begin{eqnarray}
\begin{array}{c}
H = -\frac{1}{2} \displaystyle\sum_{l \ne m} J_{lm} \vec e_l \cdot \vec e_m
\approx -\displaystyle\sum_{k=1}^6 J_k \left( \frac{1}{2}\displaystyle \sum_{l \ne m}^\text{$k$-th n.n.} \vec e_l \cdot \vec e_m \right)%
\end{array}
\end{eqnarray}
Here, $\vec e_l$ is a unit vetor that point to the direction of the spin at site $l$ and $J_{lm}$ the effective coupling constant between site $l$ and $m$. The factor of $1/2$ removes double counting. $J_k$ is the average of $J_{lm}$ taken over the $k$-th nearest-neighbor spin pairs. Note that the magnitude of spin (ideally $S=1/2$ for Cu$^{2+}$ ions) is renormalized in $J_k$ and that the summation over $k$ terminates at 6 as we consider up to $J_6$ (see Fig.~\ref{crystalmag}b).

\noindent{{\bf Data availability.}}
The data that support the findings of this study are available from the corresponding author on request.

\newpage
\noindent{}\\
\noindent{{\bf References}}

\begin{addendum}
\item[Acknowledgments]
\noindent{}\\
We thank Y. Kato and Y. Motome for helpful discussions and H. Tada for specific heat measurements. We are grateful for the initial neutron diffraction measurements by L. Keller. Experiments at the ISIS Pulsed Neutron and Muon Source were supported by a beamtime allocation from the Science and Technology Facilities Council. Neutron scattering experiments were carried out at the continuous spallation neutron source SINQ at the Paul Scherrer Institut at Villigen PSI (Switzerland). This work was partially supported by JSPS KAKENHI Grants 26610103, 16K05449, 26800186, and 24244058, by European Research Council grant CONQUEST, and by the Swiss National Science Foundation and its Sinergia network MPBH.

\item[Author contributions]
\noindent{}\\
K.K. and T.K. conceived the project. K.K., H.M.R., and T.K. coordinated experiments.
K.K. and M.S. performed crystal growth and magnetization, dielectric constant, and specific heat measurements.
T.N. and Y.N. arranged dielectric measurements in a high magnetic field.
P.B., H.M.R., G.S.T., J.M., T.F., P.M., D.D.K., and R.D.J. collected and analyzed neutron scattering data.
M.T. and K.Y. performed density functional theory calculations. K.Y. performed Landau analysis.
K.K., P.B., H.M.R., M.T., K.Y., and T.K wrote the paper.

\item[Additional information]
\noindent{}\\
Supplementary information is available in the online version of the paper. Reprints and permissions information is available online at www.nature.com/reprints.\\
Correspondence and requests for materials should be addressed to K.K.

\item[Competing financial interests]
\noindent{}\\
The authors declare no competing financial interests.

\end{addendum}

%Figure

\begin{thebibliography}{10}
\expandafter\ifx\csname url\endcsname\relax
  \def\url#1{\texttt{#1}}\fi
\expandafter\ifx\csname urlprefix\endcsname\relax\def\urlprefix{URL }\fi
\providecommand{\bibinfo}[2]{#2}
\providecommand{\eprint}[2][]{\url{#2}}

\bibitem{Greedan2001}
\bibinfo{author}{Greedan, J.~E.}
\newblock \bibinfo{title}{Geometrically frustrated magnetic materials}.
\newblock \emph{\bibinfo{journal}{J. Mater. Chem.}}
  \textbf{\bibinfo{volume}{11}}, \bibinfo{pages}{37--53}
  (\bibinfo{year}{2001}).

\bibitem{Kimura2003}
\bibinfo{author}{Kimura, T.} \emph{et~al.}
\newblock \bibinfo{title}{Magnetic control of ferroelectric polarization}.
\newblock \emph{\bibinfo{journal}{Nature}} \textbf{\bibinfo{volume}{426}},
  \bibinfo{pages}{55--58} (\bibinfo{year}{2003}).

\bibitem{Katsura2005}
\bibinfo{author}{Katsura, H.}, \bibinfo{author}{Nagaosa, N.} \&
  \bibinfo{author}{Balatsky, A.~V.}
\newblock \bibinfo{title}{Spin current and magnetoelectric effect in
  noncollinear magnets}.
\newblock \emph{\bibinfo{journal}{Phys. Rev. Lett.}}
  \textbf{\bibinfo{volume}{95}}, \bibinfo{pages}{057205}
  (\bibinfo{year}{2005}).

\bibitem{Mostovoy2006}
\bibinfo{author}{Mostovoy, M.}
\newblock \bibinfo{title}{Ferroelectricity in spiral magnets}.
\newblock \emph{\bibinfo{journal}{Phys. Rev. Lett.}}
  \textbf{\bibinfo{volume}{96}}, \bibinfo{pages}{067601}
  (\bibinfo{year}{2006}).

\bibitem{Sergienko2006}
\bibinfo{author}{Sergienko, I.~A.} \& \bibinfo{author}{Dagotto, E.}
\newblock \bibinfo{title}{{Role of the Dzyaloshinskii-Moriya interaction in
  multiferroic perovskites}}.
\newblock \emph{\bibinfo{journal}{Phys. Rev. B}} \textbf{\bibinfo{volume}{73}},
  \bibinfo{pages}{094434} (\bibinfo{year}{2006}).

\bibitem{Astrov1960}
\bibinfo{author}{Astrov, D.~N.}
\newblock \bibinfo{title}{The magnetoelectric effect in antiferromagnetics}.
\newblock \emph{\bibinfo{journal}{Sov. Phys. JETP}}
  \textbf{\bibinfo{volume}{11}}, \bibinfo{pages}{708--709}
  (\bibinfo{year}{1960}).

\bibitem{Schmid2001}
\bibinfo{author}{Schmid, H.}
\newblock \bibinfo{title}{On ferrotoroidics and electrotoroidic,
  magnetotoroidic and piezotoroidic effects}.
\newblock \emph{\bibinfo{journal}{Ferroelectrics}}
  \textbf{\bibinfo{volume}{252}}, \bibinfo{pages}{41--50}
  (\bibinfo{year}{2001}).

\bibitem{Fiebig2005}
\bibinfo{author}{Fiebig, M.}
\newblock \bibinfo{title}{Revival of the magnetoelectric effect}.
\newblock \emph{\bibinfo{journal}{J. Phys. D}} \textbf{\bibinfo{volume}{38}},
  \bibinfo{pages}{R123--R152} (\bibinfo{year}{2005}).

\bibitem{VanAken2007}
\bibinfo{author}{Aken, B.~V.}, \bibinfo{author}{Rivera, J.-P.},
  \bibinfo{author}{Schmid, H.} \& \bibinfo{author}{Fiebig, M.}
\newblock \bibinfo{title}{Observation of ferrotoroidic domains}.
\newblock \emph{\bibinfo{journal}{Nature}} \textbf{\bibinfo{volume}{449}},
  \bibinfo{pages}{702--705} (\bibinfo{year}{2007}).

\bibitem{Spaldin2008}
\bibinfo{author}{Spaldin, N.~A.}, \bibinfo{author}{Fiebig, M.} \&
  \bibinfo{author}{Mostovoy, M.}
\newblock \bibinfo{title}{The toroidal moment in condensed-matter physics and
  its relation to the magnetoelectric effect}.
\newblock \emph{\bibinfo{journal}{J. Phys.: Condens. Matter}}
  \textbf{\bibinfo{volume}{20}}, \bibinfo{pages}{434203}
  (\bibinfo{year}{2008}).

\bibitem{Yamaguchi2013}
\bibinfo{author}{Yamaguchi, Y.} \& \bibinfo{author}{Kimura, T.}
\newblock \bibinfo{title}{Magnetoelectric control of frozen state in a toroidal
  glass}.
\newblock \emph{\bibinfo{journal}{Nat. Commun.}} \textbf{\bibinfo{volume}{4}},
  \bibinfo{pages}{2063} (\bibinfo{year}{2013}).

\bibitem{Zimmermann2014}
\bibinfo{author}{Zimmermann, A.~S.}, \bibinfo{author}{Meier, D.} \&
  \bibinfo{author}{Fiebig, M.}
\newblock \bibinfo{title}{Ferroic nature of magnetic toroidal order}.
\newblock \emph{\bibinfo{journal}{Nat. Commun.}} \textbf{\bibinfo{volume}{5}},
  \bibinfo{pages}{4796} (\bibinfo{year}{2014}).

\bibitem{Dubovik1990}
\bibinfo{author}{Dubovik, V.~M.} \& \bibinfo{author}{Tugushev, V.~V.}
\newblock \bibinfo{title}{Toroid moments in electrodynamics and solid-state
  physics}.
\newblock \emph{\bibinfo{journal}{Phys. Rep.}} \textbf{\bibinfo{volume}{187}},
  \bibinfo{pages}{145--202} (\bibinfo{year}{1990}).

\bibitem{Spaldin2013}
\bibinfo{author}{Spaldin, N.~A.}, \bibinfo{author}{Fechner, M.},
  \bibinfo{author}{Bousquet, E.}, \bibinfo{author}{Balatsky, A.} \&
  \bibinfo{author}{Nordstr\"om, L.}
\newblock \bibinfo{title}{Monopole-based formalism for the diagonal
  magnetoelectric response}.
\newblock \emph{\bibinfo{journal}{Phys. Rev. B}} \textbf{\bibinfo{volume}{88}},
  \bibinfo{pages}{094429} (\bibinfo{year}{2013}).

\bibitem{Delaney2009}
\bibinfo{author}{Delaney, K.~T.}, \bibinfo{author}{Mostovoy, M.} \&
  \bibinfo{author}{Spaldin, N.~A.}
\newblock \bibinfo{title}{Superexchange-driven magnetoelectricity in magnetic
  vortices}.
\newblock \emph{\bibinfo{journal}{Phys. Rev. Lett.}}
  \textbf{\bibinfo{volume}{102}}, \bibinfo{pages}{157203}
  (\bibinfo{year}{2009}).

\bibitem{Popov1998}
\bibinfo{author}{Popov, Y.~F.} \emph{et~al.}
\newblock \bibinfo{title}{Magnetoelectric effect and toroidal ordering in
  {Ga$_{2-x}$Fe$_{x}$O$_{3}$}}.
\newblock \emph{\bibinfo{journal}{JETP}} \textbf{\bibinfo{volume}{87}},
  \bibinfo{pages}{146--151} (\bibinfo{year}{1998}).

\bibitem{Ressouche2010}
\bibinfo{author}{Ressouche, E.} \emph{et~al.}
\newblock \bibinfo{title}{Magnetoelectric {MnPS$_{3}$} as a candidate for
  ferrotoroidicity}.
\newblock \emph{\bibinfo{journal}{Phys. Rev. B}} \textbf{\bibinfo{volume}{82}},
  \bibinfo{pages}{100408} (\bibinfo{year}{2010}).

\bibitem{Baum2013}
\bibinfo{author}{Baum, M.} \emph{et~al.}
\newblock \bibinfo{title}{Controlling toroidal moments by crossed electric and
  magnetic fields}.
\newblock \emph{\bibinfo{journal}{Phys. Rev. B}} \textbf{\bibinfo{volume}{88}},
  \bibinfo{pages}{024414} (\bibinfo{year}{2013}).

\bibitem{Toledano2015}
\bibinfo{author}{Tol\'edano, P.} \emph{et~al.}
\newblock \bibinfo{title}{Primary ferrotoroidicity in antiferromagnets}.
\newblock \emph{\bibinfo{journal}{Phys. Rev. B}} \textbf{\bibinfo{volume}{92}},
  \bibinfo{pages}{094431} (\bibinfo{year}{2015}).

\bibitem{Tokura1990}
\bibinfo{author}{Tokura, Y.} \& \bibinfo{author}{Arima, T.}
\newblock \bibinfo{title}{New classification method for layered copper oxide
  compounds and its application to design of new high {$T_{c}$}
  superconductors}.
\newblock \emph{\bibinfo{journal}{Jpn. J. Appl. Phys.}}
  \textbf{\bibinfo{volume}{29}}, \bibinfo{pages}{2388--2402}
  (\bibinfo{year}{1990}).

\bibitem{Tsujimoto2007}
\bibinfo{author}{Tsujimoto, Y.} \emph{et~al.}
\newblock \bibinfo{title}{Infinite-layer iron oxide with a square-planar
  coordination}.
\newblock \emph{\bibinfo{journal}{Nature}} \textbf{\bibinfo{volume}{450}},
  \bibinfo{pages}{1062--1065} (\bibinfo{year}{2007}).

\bibitem{Johnson1966}
\bibinfo{author}{Johnson, N.~W.}
\newblock \bibinfo{title}{Convex polyhedra with regular faces}.
\newblock \emph{\bibinfo{journal}{Can. J. Math.}}
  \textbf{\bibinfo{volume}{18}}, \bibinfo{pages}{169--200}
  (\bibinfo{year}{1966}).

\bibitem{KKimura2016}
\bibinfo{author}{Kimura, K.}, \bibinfo{author}{Sera, M.} \&
  \bibinfo{author}{Kimura, T.}
\newblock \bibinfo{title}{{{$\it A$}$^{2+}$} cation control of chiral domain
  formation in {{$\it A$}(TiO)Cu$_4$(PO$_4$)$_4$} ({$\it A$} = {Ba, Sr})}.
\newblock \emph{\bibinfo{journal}{Inorg. Chem.}} \textbf{\bibinfo{volume}{55}},
  \bibinfo{pages}{1002--1004} (\bibinfo{year}{2016}).

\bibitem{Prsa2009}
\bibinfo{author}{Pr\ifmmode~\check{s}\else \v{s}\fi{}a, K.} \emph{et~al.}
\newblock \bibinfo{title}{Anomalous magnetic excitations of cooperative
  tetrahedral spin clusters}.
\newblock \emph{\bibinfo{journal}{Phys. Rev. Lett.}}
  \textbf{\bibinfo{volume}{102}}, \bibinfo{pages}{177202}
  (\bibinfo{year}{2009}).

\bibitem{PBE1996}
\bibinfo{author}{Perdew, J.~P.}, \bibinfo{author}{Burke, K.} \&
  \bibinfo{author}{Ernzerhof, M.}
\newblock \bibinfo{title}{Generalized gradient approximation made simple}.
\newblock \emph{\bibinfo{journal}{Phys. Rev. Lett.}}
  \textbf{\bibinfo{volume}{77}}, \bibinfo{pages}{3865--3868}
  (\bibinfo{year}{1996}).

\bibitem{Liechtenstein1995}
\bibinfo{author}{Liechtenstein, A.~I.}, \bibinfo{author}{Anisimov, V.~I.} \&
  \bibinfo{author}{Zaanen, J.}
\newblock \bibinfo{title}{Density-functional theory and strong interactions:
  Orbital ordering in {Mott-Hubbard} insulators}.
\newblock \emph{\bibinfo{journal}{Phys. Rev. B}} \textbf{\bibinfo{volume}{52}},
  \bibinfo{pages}{R5467--R5470} (\bibinfo{year}{1995}).

\bibitem{Toth2015}
\bibinfo{author}{Toth, S.} \& \bibinfo{author}{Lake, B.}
\newblock \bibinfo{title}{{Linear spin wave theory for single-Q incommensurate
  magnetic structures}}.
\newblock \emph{\bibinfo{journal}{J. Phys.: Condens. Matter}}
  \textbf{\bibinfo{volume}{27}}, \bibinfo{pages}{166002}
  (\bibinfo{year}{2015}).

\bibitem{Iyama2013}
\bibinfo{author}{Iyama, A.} \& \bibinfo{author}{Kimura, T.}
\newblock \bibinfo{title}{Magnetoelectric hysteresis loops in
  {Cr${}_{2}$O${}_{3}$} at room temperature}.
\newblock \emph{\bibinfo{journal}{Phys. Rev. B}} \textbf{\bibinfo{volume}{87}},
  \bibinfo{pages}{180408} (\bibinfo{year}{2013}).

\bibitem{Mufti2011}
\bibinfo{author}{Mufti, N.} \emph{et~al.}
\newblock \bibinfo{title}{Magnetoelectric coupling in {MnTiO$_{3}$}}.
\newblock \emph{\bibinfo{journal}{Phys. Rev. B}} \textbf{\bibinfo{volume}{83}},
  \bibinfo{pages}{104416} (\bibinfo{year}{2011}).

\bibitem{Birss1966}
\bibinfo{author}{Birss, R.~R.}
\newblock \bibinfo{note}{{\it{Symmetry and Magnetism} \rm{(North-Holland,
  Amsterdam, 1966)}}}.

\bibitem{Landau1980}
\bibinfo{author}{Landau, L.~D.} \& \bibinfo{author}{Lifshitz, E.~M.}
\newblock \bibinfo{note}{{\it{Statistical Physics}, \rm{Part I (Pergamon,
  Oxford, 1980)}}}.

\bibitem{Yamauchi2011}
\bibinfo{author}{Yamauchi, K.}, \bibinfo{author}{Barone, P.} \&
  \bibinfo{author}{Picozzi, S.}
\newblock \bibinfo{title}{Theoretical investigation of magnetoelectric effects
  in {Ba$_{2}$CoGe$_{2}$O$_{7}$}}.
\newblock \emph{\bibinfo{journal}{Phys. Rev. B}} \textbf{\bibinfo{volume}{84}},
  \bibinfo{pages}{165137} (\bibinfo{year}{2011}).

\bibitem{Yamauchi2014}
\bibinfo{author}{Yamauchi, K.}, \bibinfo{author}{Oguchi, T.} \&
  \bibinfo{author}{Picozzi, S.}
\newblock \bibinfo{title}{Ab-initio prediction of magnetoelectricity in
  infinite-layer {CaFeO$_{2}$} and {MgFeO$_{2}$}}.
\newblock \emph{\bibinfo{journal}{J. Phys. Soc. Jpn.}}
  \textbf{\bibinfo{volume}{83}}, \bibinfo{pages}{094712}
  (\bibinfo{year}{2014}).

\bibitem{Campbell2006}
\bibinfo{author}{Campbell, B.~J.}, \bibinfo{author}{Stokes, H.~T.},
  \bibinfo{author}{Tanner, D.~E.} \& \bibinfo{author}{Hatch, D.~M.}
\newblock \bibinfo{title}{{ISODISPLACE:} a web-based tool for exploring
  structural distortions}.
\newblock \emph{\bibinfo{journal}{J. Appl. Crystallogr.}}
  \textbf{\bibinfo{volume}{39}}, \bibinfo{pages}{607--614}
  (\bibinfo{year}{2006}).

\bibitem{Aroyo2011}
\bibinfo{author}{Aroyo, M.~I.} \emph{et~al.}
\newblock \bibinfo{title}{{Crystallography online: Bilbao crystallographic
  server}}.
\newblock \emph{\bibinfo{journal}{Bulg. Chem. Commun.}}
  \textbf{\bibinfo{volume}{43}}, \bibinfo{pages}{183--197}
  (\bibinfo{year}{2011}).

\bibitem{Aroyo2006}
\bibinfo{author}{Aroyo, M.~I.} \emph{et~al.}
\newblock \bibinfo{title}{{Bilbao Crystallographic Server: I. Databases and
  crystallographic computing programs}}.
\newblock \emph{\bibinfo{journal}{Z. Kristallogr.}}
  \textbf{\bibinfo{volume}{221}}, \bibinfo{pages}{15--27}
  (\bibinfo{year}{2006}).

\bibitem{Aroyo2006b}
\bibinfo{author}{Aroyo, M.~I.}, \bibinfo{author}{Kirov, A.},
  \bibinfo{author}{Capillas, C.}, \bibinfo{author}{Perez-Mato, J.~M.} \&
  \bibinfo{author}{Wondratschek, H.}
\newblock \bibinfo{title}{{Bilbao Crystallographic Server. II. Representations
  of crystallographic point groups and space groups}}.
\newblock \emph{\bibinfo{journal}{Acta Crystallogr. A}}
  \textbf{\bibinfo{volume}{62}}, \bibinfo{pages}{115--128}
  (\bibinfo{year}{2006}).

\bibitem{Arima2007}
\bibinfo{author}{Arima, T.}
\newblock \bibinfo{title}{Ferroelectricity induced by proper-screw type
  magnetic order}.
\newblock \emph{\bibinfo{journal}{J. Phys. Soc. Jpn.}}
  \textbf{\bibinfo{volume}{76}}, \bibinfo{pages}{073702}
  (\bibinfo{year}{2007}).

\bibitem{Hwu2002}
\bibinfo{author}{Hwu, S.-J.} \emph{et~al.}
\newblock \bibinfo{title}{A new class of hybrid materials via salt inclusion:
  Novel copper({II}) arsenates {Na$_5${$\it A$}Cu$_4$(AsO$_4$)$_4$Cl$_2$ ({$\it
  A$} = Rb, Cs)} composed of alternating covalent and ionic lattices}.
\newblock \emph{\bibinfo{journal}{J. Am. Chem. Soc.}}
  \textbf{\bibinfo{volume}{124}}, \bibinfo{pages}{12404--12405}
  (\bibinfo{year}{2002}).

\bibitem{Williams2015}
\bibinfo{author}{Williams, E.~R.}, \bibinfo{author}{Marshall, K.} \&
  \bibinfo{author}{Weller, M.~T.}
\newblock \bibinfo{title}{{Copper (II) chlorofluorophosphate: a new layered
  square-net for intercalating amines}}.
\newblock \emph{\bibinfo{journal}{CrystEngComm}} \textbf{\bibinfo{volume}{17}},
  \bibinfo{pages}{160--164} (\bibinfo{year}{2015}).

\bibitem{Momma2011}
\bibinfo{author}{Momma, K.} \& \bibinfo{author}{Izumi, F.}
\newblock \bibinfo{title}{{VESTA 3 for three-dimensional visualization of
  crystal, volumetric and morphology data}}.
\newblock \emph{\bibinfo{journal}{J. Appl. Crystallogr.}}
  \textbf{\bibinfo{volume}{44}}, \bibinfo{pages}{1272--1276}
  (\bibinfo{year}{2011}).

\bibitem{chapon-wish}
\bibinfo{author}{Chapon, L.~C.} \emph{et~al.}
\newblock \bibinfo{title}{Wish: The new powder and single crystal magnetic
  diffractometer on the second target station}.
\newblock \emph{\bibinfo{journal}{Neutron News}} \textbf{\bibinfo{volume}{22}},
  \bibinfo{pages}{22--25} (\bibinfo{year}{2011}).

\bibitem{Fisher}
\bibinfo{author}{Fischer, P.}, \bibinfo{author}{Keller, L.},
  \bibinfo{author}{Schefer, J.} \& \bibinfo{author}{Kohlbrecher, J.}
\newblock \bibinfo{title}{Neutron diffraction at {SINQ}}.
\newblock \emph{\bibinfo{journal}{Neutron News}} \textbf{\bibinfo{volume}{11}},
  \bibinfo{pages}{19--21} (\bibinfo{year}{2000}).

\bibitem{fullprof}
\bibinfo{author}{Rodriguez-Carvajal, J.}
\newblock \bibinfo{title}{Recent advances in magnetic structure determination
  by neutron powder diffraction}.
\newblock \emph{\bibinfo{journal}{Physica B}} \textbf{\bibinfo{volume}{192}},
  \bibinfo{pages}{55--69} (\bibinfo{year}{1993}).

\bibitem{janssen}
\bibinfo{author}{Jan{$\upbeta$}en, S.}, \bibinfo{author}{Mesot, J.},
  \bibinfo{author}{Holitzner, L.}, \bibinfo{author}{Furrer, A.} \&
  \bibinfo{author}{Hempelmann, R.}
\newblock \bibinfo{title}{Focus: a hybrid {TOF}-spectrometer at {SINQ}}.
\newblock \emph{\bibinfo{journal}{Physica B}}
  \textbf{\bibinfo{volume}{234-236}}, \bibinfo{pages}{1174--1176}
  (\bibinfo{year}{1997}).

\bibitem{Kresse1993}
\bibinfo{author}{Kresse, G.} \& \bibinfo{author}{Hafner, J.}
\newblock \bibinfo{title}{\textit{Ab initio} molecular dynamics for liquid
  metals}.
\newblock \emph{\bibinfo{journal}{Phys. Rev. B}} \textbf{\bibinfo{volume}{47}},
  \bibinfo{pages}{558--561} (\bibinfo{year}{1993}).

\bibitem{Kresse1994}
\bibinfo{author}{Kresse, G.} \& \bibinfo{author}{Hafner, J.}
\newblock \bibinfo{title}{\textit{Ab initio} molecular-dynamics simulation of
  the liquid-metal--amorphous-semiconductor transition in germanium}.
\newblock \emph{\bibinfo{journal}{Phys. Rev. B}} \textbf{\bibinfo{volume}{49}},
  \bibinfo{pages}{14251--14269} (\bibinfo{year}{1994}).

\bibitem{Kresse1996}
\bibinfo{author}{Kresse, G.} \& \bibinfo{author}{Furthm\"uller, J.}
\newblock \bibinfo{title}{Efficient iterative schemes for \textit{ab initio}
  total-energy calculations using a plane-wave basis set}.
\newblock \emph{\bibinfo{journal}{Phys. Rev. B}} \textbf{\bibinfo{volume}{54}},
  \bibinfo{pages}{11169--11186} (\bibinfo{year}{1996}).

\bibitem{Kresse1996_cms}
\bibinfo{author}{Kresse, G.} \& \bibinfo{author}{Furthm\"uller, J.}
\newblock \bibinfo{title}{Efficiency of \textit{ab initio} total energy
  calculations for metals and semiconductors using a plane-wave basis set}.
\newblock \emph{\bibinfo{journal}{Comput. Mat. Sci.}}
  \textbf{\bibinfo{volume}{6}}, \bibinfo{pages}{15--50} (\bibinfo{year}{1996}).

\bibitem{PAW}
\bibinfo{author}{Bl\"ochl, P.~E.}
\newblock \bibinfo{title}{Projector augmented-wave method}.
\newblock \emph{\bibinfo{journal}{Phys. Rev. B}} \textbf{\bibinfo{volume}{50}},
  \bibinfo{pages}{17953--17979} (\bibinfo{year}{1994}).

\bibitem{Kresse1999}
\bibinfo{author}{Kresse, G.} \& \bibinfo{author}{Joubert, D.}
\newblock \bibinfo{title}{From ultrasoft pseudopotentials to the projector
  augmented-wave method}.
\newblock \emph{\bibinfo{journal}{Phys. Rev. B}} \textbf{\bibinfo{volume}{59}},
  \bibinfo{pages}{1758--1775} (\bibinfo{year}{1999}).

\end{thebibliography}
\end{document}